\documentclass[1p]{elsarticle}

\usepackage{lineno,hyperref}
\modulolinenumbers[5]

\journal{Journal of Geometry and Physics}

\usepackage{graphicx}
\usepackage[inline]{enumitem}
\usepackage{subcaption} 
\usepackage{colortbl}
\usepackage{float}

\usepackage{natbib}
\usepackage{bibentry}

\usepackage{multirow}
\usepackage{hyphenat}
\usepackage[ruled, vlined, linesnumbered]{algorithm2e}
\usepackage{rotating} 
\usepackage{adjustbox}%
\usepackage[11pt]{moresize}
\usepackage{gensymb} 

\usepackage{amsthm}
\usepackage{amsmath}
\usepackage{amssymb}
\usepackage{amsfonts}

\usepackage{stmaryrd} 

\usepackage[normalem]{ulem}

\usepackage{csquotes}

\usepackage{mathrsfs}

\newtheorem{theorem}{Theorem}[section]

\newtheorem{proposition}[theorem]{Proposition}

\theoremstyle{definition}

\newtheorem{remark}[theorem]{Remark}

\hypersetup{hidelinks}

\begin{document}

\begin{frontmatter}

\title{Port-Hamiltonian Modeling of Ideal Fluid Flow: \\Part II. Compressible and Incompressible Flow}

\author{Ramy Rashad\textsuperscript{1,*}\corref{RAM}, Federico Califano\textsuperscript{1}, Frederic P. Schuller\textsuperscript{2}, Stefano Stramigioli\textsuperscript{1}}
\address{\textsuperscript{1} Robotics and Mechatronics Department, University of Twente, The Netherlands}
\address{\textsuperscript{2} Department of Applied Mathematics, University of Twente, The Netherlands}
\cortext[RAM]{Corresponding author. Email: r.a.m.rashadhashem@utwente.nl}

%
%

\begin{abstract}
Part I of this paper presented a systematic derivation of the Stokes Dirac structure underlying the port-Hamiltonian model of ideal fluid flow on Riemannian manifolds.
Starting from the group of diffeomorphisms as a configuration space for the fluid, the Stokes Dirac structure is derived by Poisson reduction and then augmented by boundary ports and distributed ports. The additional boundary ports have been shown to appear naturally as surface terms in the pairings of dual maps, always neglected in standard Hamiltonian theory.
The port-Hamiltonian model presented in Part I corresponded only to the kinetic energy of the fluid and how its energy variables evolve such that the energy is conserved.

In Part II, we utilize the distributed port of the kinetic energy port-Hamiltonian system for representing a number of fluid-dynamical systems.
By adding internal energy we model compressible flow, both adiabatic and isentropic, and by adding constraint forces we model incompressible flow.
The key tools used are the interconnection maps relating the dynamics of fluid motion to the dynamics of advected quantities.
\end{abstract}


\end{frontmatter}


\newcommand{\half}{\frac{1}{2}}

\newcommand{\B}[1]{\boldsymbol{#1}}
\newcommand{\bb}[1]{\mathbb{#1}}
\newcommand{\cl}[1]{\mathcal{#1}}
\newcommand{\Rn}{\bb{R}^n}

\newcommand{\TwoTwoMat}[4]{
\begin{pmatrix}
#1 & #2 \\
#3 & #4
\end{pmatrix}}
\newcommand{\TwoVec}[2]{
\begin{pmatrix}
#1\\
#2 
\end{pmatrix}}

\newcommand{\ThrVec}[3]{
\begin{pmatrix}
#1\\
#2\\
#3
\end{pmatrix}}

\newcommand{\map}[3]{{#1}:{#2}\rightarrow {#3}}
\newcommand{\fullmap}[5]{
\begin{split}
#1 : {#2} &\rightarrow {#3}\\
	{#4} &\mapsto {#5}
\end{split}
}
\newcommand{\pair}[2]{\left\langle \left.  #1  \right|  #2 \right\rangle}

\newcommand{\blue}{\color{blue}}

\newcommand{\pH}{port-Hamiltonian }

\newcommand{\gothg}{\mathfrak{g}}
\newcommand{\gothgV}{\mathfrak{g^v}}
\newcommand{\gothgstar}{\mathfrak{g}^{*}}

\newcommand{\extd}{\textrm{d}}

\newcommand{\goths}{\mathfrak{s}}
\newcommand{\gothsStar}{\mathfrak{s}^{*}}

\newcommand{\secTanBdl}[1]{\Gamma(T#1)}

\newcommand{\volF}{\mu_{\normalfont\text{vol}}}

\newcommand{\abar}{{\bar{a}}}


\newcommand{\LieD}[2]{\cl{L}_{#1}#2}
\newcommand{\dt}{\frac{d}{dt}}

\newcommand{\dtLine}[1]{\frac{d}{dt}\bigg|_{#1}}
\newcommand{\depsLine}[1]{\frac{d}{d \epsilon}\bigg|_{#1}}

\newcommand{\pt}{\frac{\partial}{\partial t}}
\newcommand{\JLBrack}[3]{\llbracket{ #2},{#3} \rrbracket_{#1}}

\newcommand{\inner}[2]{\left \langle #1 , #2 \right\rangle}

\newcommand{\Lbrack}[2]{\left[#1,#2 \right]}
\newcommand{\LbrG}[2]{\Lbrack{#1}{#2}_{\gothg}}
\newcommand{\LbrS}[2]{\Lbrack{#1}{#2}_{\goths}}

\newcommand{\Pbrack}[2]{\left \{#1,#2 \right\}}

\newcommand{\parXi}[1][i]{\frac{\partial}{\partial x^{#1}}}
\newcommand{\dXi}[1][i]{\extd{ x^{#1}}}

\newcommand{\varD}[2]{\frac{\delta {#1}}{\delta {#2}}}

\newcommand{\adS}[1]{\boldsymbol{ad}_{#1}}
\newcommand{\adSdual}[1]{\boldsymbol{ad}^*_{#1}}

\newcommand{\spKForm}[2]{\Omega^{#1}(#2)}
\newcommand{\spVecF}[1]{\Gamma(T#1)}
\newcommand{\spVecX}[2]{\mathfrak{X}_{#2}(#1)}
\newcommand{\spKFormM}[1]{\spKForm{#1}{M}}
\newcommand{\spKFormMbc}[2]{\Omega^{#1}_{#2}(M)}
\newcommand{\spVecM}{\spVecX{M}{}}
\newcommand{\spVecMt}{\spVecX{M}{T}}
\newcommand{\spFn}[1]{C^\infty(#1)}

\newcommand{\vMeas}[1]{\textbf{v}(#1)}
\newcommand{\mMeas}[1]{\textbf{m}(#1)}
\newcommand{\mForm}{\mu_t}
\newcommand{\divr}[1]{\text{\normalfont div}(#1)}
\newcommand{\sdiff}{D_\mu(M)}
\newcommand{\RVF}[1]{\cl{R}(#1)}
\newcommand{\connect}[1]{\stackrel{\bb{#1}}{\nabla}}
\newcommand{\adIflat}[2]{ad_{#1}^*(\bb{I}^\flat {#2})}
\newcommand{\vf}{{\tilde{v}}}
\newcommand{\uf}{\tilde{u}}
\newcommand{\omV}{{\hat{\omega}}}

\newcommand{\mNmOne}{(-1)^{n-1} }

\newcommand{\effOne}{\delta_{\mu}H}
\newcommand{\effTwo}{\delta_{\alpha}H}
\newcommand{\bound}[1]{{#1}\rvert_{ \partial M}}
\newcommand{\effOneB}{\bound{\delta_{\mu}H}}
\newcommand{\effTwoB}{\bound{\delta_{\alpha}H}}

\newcommand{\gtInv}{g^{-1}_t}
\newcommand{\diffG}{\mathcal{D}(M)}

\newcommand{\parL}[1]{\frac{\delta l}{\delta {#1}}}

\newcommand{\Mbound}{\partial M}

\newcommand{\Ltwo}{\cl{L}_2}
\newcommand{\Mflat}{\bb{M}^\flat}
\newcommand{\Msharp}{\bb{M}^\sharp}

\newcommand{\state}{(\alpha,\mu)}

\tableofcontents


%
\newcommand{\mapR}{\tilde{\varphi}_a}
\newcommand{\mapRdual}{\tilde{\varphi}_a^*}
\newcommand{\surfPhiDTA}[1]{\eta_{\tilde{\varphi}_{#1}}}

\newcommand{\eSV}{{e}_{sk}}
\newcommand{\fSV}{{f}_{sk}}

\newcommand{\eMu}{\delta_\mu H_k}
\newcommand{\fMu}{\dot{\mu}}
\newcommand{\fstr}{\text{f}_\text{s}}
\newcommand{\eBoud}{e_{\partial k}}
\newcommand{\fBoud}{f_{\partial k}}

\newcommand{\eVt}{\delta_{\vf} H_k}
\newcommand{\fVt}{\dot{\vf}}
\newcommand{\eXt}{(e_{\vf},e_{\tilde{\mu}})}

\newcommand{\DkT}{\cl{\tilde{D}}_k}
\newcommand{\stateV}{(\vf,{\mu})}

\section{Introduction}
In Part II of this paper, we present the port-Hamiltonian models of a number of fluid dynamical systems on general Riemannian manifolds.
We start from the velocity representation of the port-Hamiltonian model describing the evolution of the kinetic energy of the fluid.
The explicit dynamical equations, derived in Part I and repeated here for the reader's convenience, were given in terms of the kinetic energy state variable $x_k = (\vf,\mu) \in \cl{X} = \gothsStar$ by
\begin{align}
\TwoVec{\fVt}{\fMu} =& \TwoVec{-\extd (\eMu) -  \iota_{v} \extd \vf } {-\extd (\eVt)} + \TwoVec{ \frac{1}{*\mu}}{0}\fstr, \label{eq:pH_kinetic_sys_new_1}\\
\omega_v =& \begin{pmatrix}  \frac{1}{*\mu} &  0\end{pmatrix} \TwoVec{\eVt}{\eMu}, \label{eq:pH_kinetic_sys_new_2}\\
H_k(x_k) =& H_k\stateV = \int_M \half (*\mu) \vf \wedge * \vf.\label{eq:pH_kinetic_Ham_new}
\end{align}
The variational derivatives $\delta_{\vf} H_k \in \spKFormM{n-1}$ and $\delta_{\mu} H_k \in \spKFormM{0}$ with respect to the states $\vf \in \gothgstar = \spKFormM{1} $ and $\mu \in V^* = \spKFormM{n}$, respectively, are given by
\begin{equation}\label{eq:varDeriv_stateV}
\delta_{\vf} H_k  = (*\mu) * \vf = \iota_v \mu, \qquad\qquad \delta_{\mu} H_k = \half \iota_v \vf.
\end{equation}

It was shown in Part I, that the port-Hamiltonian system (\ref{eq:pH_kinetic_Ham_new}) can be represented by a kinetic energy storage port in addition to two open ports that can be interconnected to other systems. Namely, the boundary port $(\eBoud,\fBoud) = (\bound{\half \iota_v\vf}, -\bound{\iota_v\mu}) \in \spKForm{0}{\Mbound} \times \spKForm{n-1}{\Mbound}$ and the distributed port $(e_d,f_d) = (\fstr,\omega_v) \in \gothgstar\times \gothg=\spKFormM{1} \times \spKFormM{n-1}$.
The kinetic energy Hamiltonian $H_k$ satisfies the power balance
\begin{equation}
\label{eq:new_power_balance}
\dot{H}_k=  \int_{\Mbound} \eBoud \wedge \fBoud   + \int_M e_d \wedge f_d,
\end{equation} 
stating that the change in kinetic energy is due to the sum of added power due to mass inflow through the boundary port or due to the stress forces through distributed port.

The three ports of the port-Hamiltonian system (\ref{eq:pH_kinetic_Ham_new}) were connected via the underlying Stokes-Dirac structure given by
\begin{equation}\label{eq:SDS_kinetic_system_new}
\begin{split}
\DkT = \{ (\fSV, &\fBoud, f_d,\eSV,\eBoud,e_d) \in \cl{B}_k | \\
			 \TwoVec{f_\vf}{f_\mu} &= \TwoVec{\extd e_\mu +  \frac{1}{*\mu} \iota_{\hat{e}_{\vf}} \extd \vf }{\extd e_\vf }- \TwoVec{\frac{1}{*\mu}}{0 }e_d,\\
			f_d&= \begin{pmatrix} \frac{1}{*\mu} &  0\end{pmatrix} \TwoVec{e_\vf}{e_\mu},\\ 
			 \TwoVec{\eBoud}{\fBoud}  &=  \TwoTwoMat{0}{1}{-1}{0}\TwoVec{\bound{e_\vf}}{\bound{e_\mu}}\}.
\end{split}
\end{equation}

So far we have deliberately considered only storage of kinetic energy in the fluid system and neglected potential/internal energy.
In part II of this paper, we discuss how the port-Hamiltonian system  (\ref{eq:pH_kinetic_sys_new_1}-\ref{eq:pH_kinetic_Ham_new}) will be extended to represent physically meaningful fluid dynamic systems. Namely, isentropic and adiabatic compressible flow as well as incompressible flow.

\begin{table}
\centering
\renewcommand{\arraystretch}{1.5}
\begin{tabular}{m{2.3cm} m{2cm} c c c}
\hline 
Advected Quantity $(a)$ &  Advection Space $(V^*)$ & $\mapR(\omega)$ &  $\mapRdual(\abar) $& $\surfPhiDTA{\omega}(a,\abar)$ \\\hline
Mass form $(\mu)$ &  $\spKFormM{n}$ & $\LieD{\omV}{\mu}$ & $ -(*\mu) \extd \bar{\mu}$& $-(*\mu)\omega \wedge \bar{\mu}$\\
Entropy $(s)$ & $\spKFormM{0}$ & $\LieD{\omV}{s}$ & $(*\bar{s}) \extd s $& $0$\\\hline
\end{tabular}
\caption{Summary of interconnection maps for the advected quantities: mass form $\mu \in \spKFormM{n}$ and entropy function $s \in \spKFormM{0}$. Their associated dual elements are denoted by $\bar{\mu} \in \spKFormM{0}$ and $\bar{s} \in \spKFormM{n}$, respectively.}
\label{table:intercon_maps}
\end{table}

The key tool that will allow relating the advected quantities dynamics defined on $V^* \times V$ to the distributed port defined on $\gothg\times \gothgstar$ will be the interconnection maps
\begin{equation}\label{eq:def_Rmap_and_dual}
\fullmap{\mapR}{\gothg}{V^*}{\omega}{\mapR(\omega) := \LieD{\omV}{a},} ,\qquad
\fullmap{\mapRdual}{V}{\gothgstar}{\abar}{\mapRdual(\abar),}
\end{equation}
introduced in Prop. 3.1 of Part I, and summarized in Table \ref{table:intercon_maps}.
The primary map $\mapR$ and its dual $\mapRdual$ are related to each other by
\begin{equation}\label{eq:pair_Rmap}
\pair{\mapRdual(\abar)}{\omega}_\gothg = \pair{\abar}{\mapR(\omega)}_{V^*}+ \int_{\Mbound} \bound{\surfPhiDTA{\omega}(a,\abar)}.
\end{equation}

The remainder of this paper is organized as follows.
In Sec. 2, we demonstrate how the distributed port will be utilized to add internal energy for developing port-Hamiltonian models for isentropic and adiabatic compressible flow.
Then, we follow the same procedure to add constraint forces to develop a port-Hamiltonian model for incompressible flow in Sec. 3.
Finally, we conclude this article in Sec. 4.

\section{Port-Hamiltonian Modeling of Compressible Flow}

\newcommand{\parU}[1]{\frac{\partial U}{\partial #1}}

The distributed force $\fstr$ present in the model (\ref{eq:pH_kinetic_sys_new_1}) originates physically from the random motion and collisions of the molecules that comprise the fluid.
The force $\fstr$ is defined through averaging the momentum transfer of a large group of molecules over a short time scale, compared to the macroscopic motion of the fluid encoded by the vector field $v$.
Thus, the transfer of momentum on the microscopic scale is equivalent to the continuous force $\fstr$ acting at each point in the spatial domain $M$ at the macroscopic scale.

There are two types of basic forces due to the microscopic motion of the fluid; pressure forces and viscous friction forces.
Both pressure and viscous forces are \textit{forces of stress}.
In this work, we will consider ideal flow, and thus model pressure forces only, while modeling the viscous forces is an issue of future work.

The molecular kinetic and vibration energy is encoded, at the macroscopic scale, as a continuous function $\bar{U} := \rho U \in \spFn{M}$, called the internal energy density, where $\rho$ is the mass density function, and $U$ is the specific internal energy (i.e. per unit mass).
The first law of thermodynamics states that the internal energy $\bar{U}$ is conserved only if the system is isolated, i.e. does not interact with its surrounding.
The internal energy of a system changes if there is transfer of mass and heat to or from the system, and by work done on or by the system.

The specific internal energy $U(\nu,s)$ depends on the fluid's specific volume $\nu = 1/\rho \in \spFn{M}$ and the fluid's specific entropy $s \in \spFn{M}$.
The differential of the internal energy $\extd U \in \spKFormM{1}$ is given by the famous Gibbs equation
\begin{equation}\label{eq:Gibbs_1}
\extd U(\nu,s) = -p \extd \nu + T \extd s,
\end{equation}
where $T \extd s$ corresponds to the heat exchanged per unit mass, and $p \extd \nu$ corresponds to the mechanical work done by the fluid system due to pressure forces.
Note that from (\ref{eq:Gibbs_1}) we have that
\begin{equation}\label{eq:components_one_form_dU}
\parU{\nu} = -p, \qquad \parU{s} = T,
\end{equation}
are the components of the one-form $\extd U$.

A more convenient form of Gibbs equation is given by
\begin{equation}
\extd U(\rho,s) = \frac{p}{\rho^2} \extd \rho + T \extd s,
\end{equation}
which follows from the chain rule $\parU{\rho} = \parU{\nu} \frac{\partial \nu}{\partial \rho} = p/{\rho^2}$.
The relation between the pressure $p$, specific internal energy $U$ and the mass density, given by 
\begin{equation}\label{eq:eqn_state}
p = \rho^2 \parU{\rho},
\end{equation}
is known as the \textit{equation of state} of the fluid, which should be specified for a choice of fluid.

Another useful thermodynamic variable is the specific enthalpy $h\in \spFn{M}$, related to the internal energy by the Legendre transformation.
The enthalpy can be expressed as
\begin{equation}\label{eq:enthalpy_def}
h = U + \frac{p}{\rho} = U + \rho \parU{\rho} = \frac{\partial}{\partial \rho}(\rho U),
\end{equation}
where the second equality follows from (\ref{eq:eqn_state}), while the last equality follows from the chain rule.
In terms of the enthalpy, the Gibbs equation becomes
\begin{equation}\label{eq:Gibbs_eqn_2}
\extd h(\rho,s) = \frac{\extd p}{\rho} +  T \extd s.
\end{equation}

In general, the specific entropy function $s_t\in \spFn{M}$ is not an advected quantity.
However, in the case of \textit{adiabatic compressible flow}, $s_t$ is advected with the flow.
Thus, it satisfies
\begin{equation}\label{eq:entropy_continuity}
\pt s_t + \LieD{v}{s_t} = 0.
\end{equation}

A consequence of the entropy conservation (\ref{eq:entropy_continuity}), is that if the entropy is homogeneous in space initially (i.e. $s_0(x_0) =s_0$ is constant) then it remains constant in space for all time (i.e. $s_t(x) =s_0, \forall t>0,x\in M$), and thus $\extd s = 0$.
In such case, the compressible flow is called isentropic and the specific internal energy $U(\rho)$ depends on the density $\rho$ only.
Therefore, the two forms of Gibbs equations (\ref{eq:Gibbs_1}) and (\ref{eq:Gibbs_eqn_2}) become
\begin{equation}
\extd U(\rho) = \frac{p}{\rho^2} \extd \rho, \qquad \qquad  \extd h(\rho) = \frac{\extd p}{\rho}.
\end{equation}

Next, we show how to systematically represent the pressure forces using the distributed force $\fstr$ in (\ref{eq:pH_kinetic_sys_new_1}) acting on an infinitesimal fluid element at a point in $M$.
For ease of presentation, we first consider the case of isentropic flow, followed by the slightly more general case of adiabatic flow, describing a fluid with no irreversible thermodynamic phenomena, but in which the advected entropy function $s_t$, might not be constant in space.


\subsection{Isentropic Compressible Flow}
\newcommand{\eMuI}{\delta_{\mu} H_i}
\newcommand{\mapTdualM}[1]{\frac{1}{*\mu}\tilde{\varphi}^*_{#1}}
\newcommand{\mapTM}[1]{\frac{1}{*\mu}\tilde{\varphi}_{#1}}
\newcommand{\mapTdual}[1]{\tilde{\varphi}^*_{#1}}
\newcommand{\mapT}[1]{\tilde{\varphi}_{#1}}
\newcommand{\eBoudT}{{e}_\partial}
\newcommand{\fBoudT}{{f}_\partial}
\newcommand{\Dis}{\cl{D}_{is}}

In the port-Hamiltonian system (\ref{eq:pH_kinetic_sys_new_1}) in which only kinetic energy is present, the distributed port $(e_d,f_d)$ could be used to add storage of internal energy of the fluid.
The storage of the fluid's total internal energy $H_i(\mu)$ is represented by a storage element with state manifold $\cl{X}_i = V^* = \spKFormM{n}$ and its corresponding state variable $x_i=\mu$ being the mass form.
The internal energy Hamiltonian $\map{H_i}{\cl{X}_i}{\bb{R}}$ is given by
\begin{equation}\label{eq:Ham_internal}
H_i(\mu) = \int_M U(*\mu) \mu,
\end{equation}
where $U(*\mu) = U(\rho)$ is the specific internal energy introduced earlier.

The effort and flow variables of the internal energy storage element are given by
\begin{equation}\label{eq:port_variables_internal}
\eMuI \in T_{x_i}^* \cl{X}_i \cong V = \spKFormM{0}, \qquad \fMu \in T_{x_i} \cl{X}_i \cong V^* = \spKFormM{n},
\end{equation}
where $\eMuI$ is given by the following result.

\begin{proposition}\label{prop:varHi_mu}
The variational derivative of the Hamiltonian functional $\map{H_i}{V^*}{\bb{R}}$ in (\ref{eq:Ham_internal}) with respect to $\mu \in V^* = \spKFormM{n}$, denoted by $\eMuI \in V = \spKFormM{0}$,  is equal to the enthalpy function (\ref{eq:enthalpy_def}):
\begin{equation}
\eMuI = h \in \spFn{M}.
\end{equation}
\end{proposition}
\begin{proof}
\newcommand{\muEps}{\mu_\epsilon}
\newcommand{\rhoEps}{\rho_\epsilon}
\newcommand{\dEps}{\depsLine{\epsilon = 0}}
The variational derivative $\eMuI \in \spFn{M}$ is defined implicitly as the function satisfying
\begin{equation}\label{eq:VarHi_1}
\pair{\eMuI}{\delta \mu}_{V^*} = \dEps H_i (\mu + \epsilon \delta \mu),
\end{equation}
for any $\epsilon \in \bb{R}$ and $\delta \mu \in \spKFormM{n}$.
For notational simplicity, we introduce $\muEps := \mu +\epsilon \delta \mu \in \spKFormM{n}, \rho := *\mu \in \spFn{M}, $ and $\delta \rho := *\delta \mu \in \spFn{M}$ .
Consequently, we have that $\rhoEps:= \rho + \epsilon \delta \rho = *\muEps$, as well as
\begin{equation}\label{eq:VarHi_2}
\dEps \muEps = \delta\mu.
\end{equation} 

Using (\ref{eq:Ham_internal}), (\ref{eq:VarHi_2}), and the Leibniz rule, we can rewrite (\ref{eq:VarHi_1}) as
\begin{align}
\int_M \eMuI \wedge \delta\mu 	&= \int_M \dEps (U(*\muEps) \wedge \muEps)
								= \int_M \dEps U(*\muEps) \wedge \muEps + U(*\muEps) \wedge \dEps \muEps,\nonumber\\
								&= \int_M \dEps U(*\muEps) \wedge \muEps + U(*\mu) \wedge \delta\mu.\label{eq:VarHi_3}
\end{align}
Since $\map{U}{\spFn{M}}{\spFn{M}}$ is a function on $\spFn{M}$, its derivative $\frac{d U}{d \rho}(\rho) \in \spFn{M}$ is defined implicitly as the function satisfying
\begin{equation}\label{eq:VarHi_4}
\frac{d U}{d \rho}(\rho) \cdot \delta\rho = \dEps U(\rhoEps).
\end{equation}
By substituting (\ref{eq:VarHi_4}) into (\ref{eq:VarHi_3}), we get
\begin{equation}\label{eq:VarHi_5}
\int_M \eMuI \wedge \delta\mu  = \int_M \frac{d U}{d \rho}(\rho) \cdot \delta\rho \wedge \mu + U(\rho) \wedge \delta\mu.
\end{equation}
Using the equality
$\delta\rho \wedge \mu = \delta\rho \cdot \rho \wedge \volF =  \rho\cdot \delta\rho \wedge \volF =\rho\cdot \wedge \delta \mu ,$
we can rewrite (\ref{eq:VarHi_5}) as
\begin{equation}
\int_M \eMuI \wedge \delta\mu  = \int_M (\frac{d U}{d \rho}(\rho) \cdot \rho + U(\rho) )\wedge \delta\mu.
\end{equation}
Therefore, using the chain rule, we have that
\begin{equation}
\eMuI = \frac{d U}{d \rho}(\rho) \cdot \rho + U(\rho) = \frac{d}{d \rho}(\rho \cdot U(\rho)),
\end{equation}
which is equal to the enthalpy as defined by (\ref{eq:enthalpy_def}).
Note that in case $U$ is a multi-variable function of $\rho$, the derivative $\frac{d U}{d \rho}$ in this proof is replaced by a partial derivative.
\end{proof}

\begin{figure}
\centering
\includegraphics[width=\textwidth]{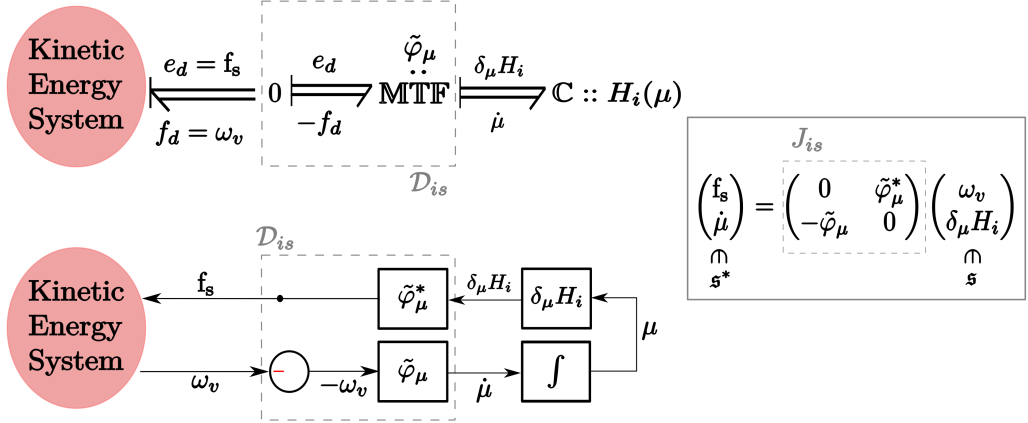}
\caption{Augmenting the kinetic energy system (\ref{eq:pH_kinetic_sys_new_1}) with the storage of internal energy through the distributed port $(e_d,f_d)$. The model corresponds to isentropic compressible flow on a manifold without boundary. The Bond graph (top) and block diagram (bottom) representations are shown.}
\label{fig:connecting_internal_energy}
\end{figure}

The Hamiltonian $H_i$ satisfies the power balance
\begin{equation}
\dot{H}_i = \pair{\eMuI}{\fMu}_{V^*}.
\end{equation}
Since the flow does not exchange heat with its surrounding, any change in the internal energy of the system is caused by the transformation of kinetic energy ({ if we assume there is no mass-flow through the boundary}).
The power incoming the internal energy storage element $\pair{\eMuI}{\fMu}_{V^*}$ is then equal to the power outcoming the distributed port of the kinetic energy subsystem, i.e.
\begin{equation}
\pair{\eMuI}{\fMu}_{V^*} = -\pair{e_d}{f_d}_\gothg = \pair{\fstr}{-\omega_v}_\gothg,
\end{equation}
as shown in Fig. \ref{fig:connecting_internal_energy}.

However, to interconnect the internal energy storage port to the kinetic energy distributed port they should be compatible.
Their incompatibility lies in the fact that $(\eMuI,\fMu) \in V\times V^*$ while $(e_d,f_d)\in \gothg\times \gothgstar$ by their definitions.
The key to connecting these two ports is related to the semi-direct product structure of $\gothg$ and $V$, and more precisely the interconnection maps (\ref{eq:def_Rmap_and_dual}) 
For simplicity, we first introduce the idea of interconnecting the two ports assuming $M$ has no boundary, then we consider the general case with the boundary port variables.

The two ports are made compatible by the use of a power-conserving transformation that relates the efforts of the two ports to each other, and relates the flows of the two ports to each other.
In the bond graph in Fig. \ref{fig:connecting_internal_energy}, the modulated transformer element $\bb{MTF}$ implements the map
\begin{equation}\label{eq:MTF_eqn}
\TwoVec{e_d}{\fMu} = \TwoTwoMat{0}{\mapTdual{\mu} }{\mapT{\mu}}{0} \TwoVec{-f_d}{\eMuI},
\end{equation}
where the map $\map{\mapT{\mu}}{\gothg}{V^*}$ and its dual $\map{\mapTdual{\mu}}{V}{\gothgstar}$ are given in Table 1 for $\mu$ as the advected parameter (i.e, $a =\mu \in V^*$).
The minus sign next to $f_d$ in (\ref{eq:MTF_eqn}) is due to the zero junction in Fig. \ref{fig:connecting_internal_energy}, used to represent the power inversion (from inflow to outflow) given by
\begin{equation}
\pair{e_d}{f_d}_\gothg = - \pair{e_d}{-f_d}_\gothg.
\end{equation}

Both the zero-junction and the  $\bb{MTF}$ combined represent a power-conserving Dirac structure {$\Dis$}, given by the image of the map $\map{J_{is}}{\goths}{\gothsStar}$, as illustrated in Fig. \ref{fig:connecting_internal_energy}.
The Dirac structure {$\Dis$} is modulated by the mass form $\mu$ (as a state of advected quantity), and its power-preserving property is clearly visible by the skew-symmetry of the map $J_{is}$.

Now we show that the previous energy-based construction correctly models compressible isentropic flow.
Consider the following equalities
\begin{align}
 \pair{\fstr}{-\omega_v}_\gothg =  \pair{\mapTdual{\mu}(\eMuI)}{-\omega_v}_\gothg 
 									= \pair{\eMuI}{\mapT{\mu}(-\omega_v)}_{V^*} = \pair{\eMuI}{\fMu}_{V^*} = \dot{H}_i,\label{eq:interconn_pairing}
\end{align}
which follows using (\ref{eq:MTF_eqn}) and the port variables definitions.
Therefore, using the expressions of $\mapT{\mu}$ and  $\mapTdual{\mu}$ in Table \ref{table:intercon_maps}, we have that 
\begin{align}
\fMu &= \mapT{\mu}(-\omega_v) = - \mapT{\mu}(\omega_v) = - \LieD{\hat{\omega}_v}{\mu} = -\LieD{v}{\mu}, \label{eq:fMu}\\
\fstr &= \mapTdual{\mu}(\eMuI) = - (*\mu) \extd(\eMuI) = - (*\mu)\extd h = - \extd p.\label{eq:fbody}
\end{align}
Therefore, (\ref{eq:fMu}) correctly represents the evolution of $\mu$ as being advected with the flow, while (\ref{eq:fbody}) correctly represents the stress forces due to pressure applied within $M$ \cite[pg. 588]{Abraham1988}.

\begin{figure}
\centering
\includegraphics[width=\textwidth]{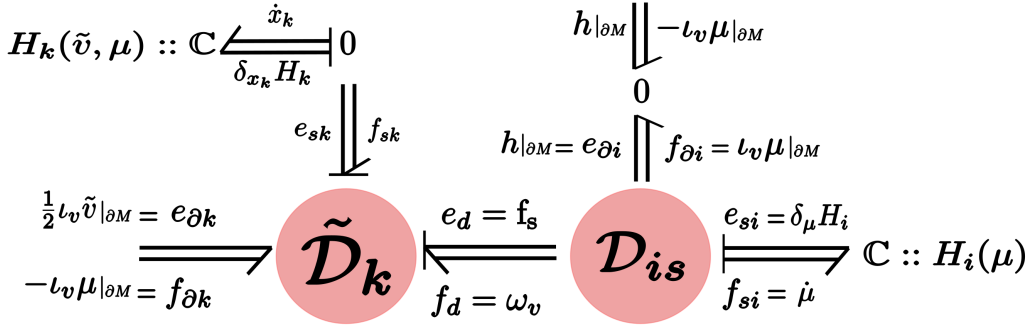}
\caption{ Augmenting the kinetic energy system (\ref{eq:pH_kinetic_sys_new_1}) with the storage of internal energy through the distributed port $(e_d,f_d)$. The model corresponds to  compressible isentropic flow on a general manifold with permeable boundary.}
\label{fig:connecting_internal_energy_with_Boundary}
\end{figure}

\newcommand{\eSint}{{e}_{si}}
\newcommand{\fSint}{{f}_{si}}
\newcommand{\eBoudInt}{e_{\partial i}}
\newcommand{\fBoudInt}{f_{\partial i}}

Now in case $M$ has a permeable boundary, the pairing equality (\ref{eq:interconn_pairing}) is no loner valid and should be augmented with a surface term $\surfPhiDTA{\mu}$ from (\ref{eq:pair_Rmap}).
{ In this case, using the expression of $\surfPhiDTA{\mu}$ in Table \ref{table:intercon_maps}, (\ref{eq:interconn_pairing}) is rewritten as
\begin{align}
 \pair{\fstr}{-\omega_v}_\gothg &=  \pair{\mapTdual{\mu}(\eMuI)}{-\omega_v}_\gothg
 									= \pair{\eMuI}{\mapT{\mu}(-\omega_v)}_{V^*} + \int_{\Mbound} -(*\mu) (-\omega_v)\bound{}\wedge\eMuI\bound{} \nonumber\\
 									& = \pair{\eMuI}{\fMu}_{V^*} + \int_{\Mbound} h\bound{}\wedge(*\mu)\omega_v\bound{}
 									= \dot{H}_i + \int_{\Mbound} h\bound{}\wedge\iota_v\mu\bound{}.\label{eq:interconn_pairing_withBound}
\end{align}
By defining the boundary port variables $\eBoudInt := h\bound{} = \eMuI\bound{}$ and $\fBoudInt := \iota_v\mu\bound{}$, the pairing equality (\ref{eq:interconn_pairing_withBound}) becomes
\begin{equation}\label{eq:power_balance_int_isent}
\pair{\fstr}{\omega_v}_\gothg + \dot{H}_i + \int_{\Mbound} \eBoudInt\wedge \fBoudInt = 0.
\end{equation}

Therefore, as shown in Fig. \ref{fig:connecting_internal_energy_with_Boundary}, the interconnection to model isentropic flow is achieved by the Dirac structure $\Dis$ given by
\begin{equation}
\begin{split}
\Dis  = \{ (\fSint, &\fBoudInt, f_d,\eSint,\eBoudInt,e_d) \in \cl{B}_{is}  | \\
			 \TwoVec{e_d}{\fSint} &= \TwoTwoMat{0}{\mapTdual{\mu} }{-\mapT{\mu}}{0} \TwoVec{f_d}{\eSint},\\
			 \TwoVec{\eBoudInt}{\fBoudInt}  &= \TwoTwoMat{1}{0}{0}{\bound{*\mu}}\TwoVec{\bound{\eSint}}{\bound{f_d}}\},\label{eq:Int_Dirac_structure}
\end{split}
\end{equation}
where the bond-space $\cl{B}_{is}= \cl{F}_{is}\times \cl{E}_{is}$ is the product space of the flow space $\cl{F}_{is}=  \spKFormM{n} \times \spKForm{n-1}{\partial M} \times \spKForm{n-1}{M} $ and the effort space $\cl{E}_{is} = \spKFormM{0} \times \spKForm{0}{\partial M} \times \spKForm{1}{M}$.
The Dirac structure (\ref{eq:Int_Dirac_structure}) is modulated by the mass form $\mu \in V^*$, and encodes the power balance
$$\pair{e_d}{f_d}_\gothg + \pair{\eSint}{\fSint}_{V^*} + \int_{\Mbound} \eBoudInt\wedge\fBoudInt = 0,$$
which is equivalent to (\ref{eq:power_balance_int_isent}) by setting the ports of $\cl{D}_{is}$ by
$$(\fMu,\bound{\iota_v\mu},\omega_v,\eMuI,\bound{h},\fstr )\in \Dis,$$
as illustrated in Fig. \ref{fig:connecting_internal_energy_with_Boundary}, and thus restoring (\ref{eq:fMu}) and (\ref{eq:fbody}).
}

In conclusion, the port-Hamiltonian model for compressible isentropic flow consists of two storage elements for kinetic and internal energy, two boundary ports $(\eBoud,\fBoud)$ and $(\eBoudInt,\fBoudInt)$ representing power through the boundary of $M$ due to mass inflow, and all the remaining power conserving elements that allow the interconnection of the aforementioned ports, shown in Fig. \ref{fig:connecting_internal_energy_with_Boundary}.

\begin{figure}
\centering
\includegraphics[width=0.6\textwidth]{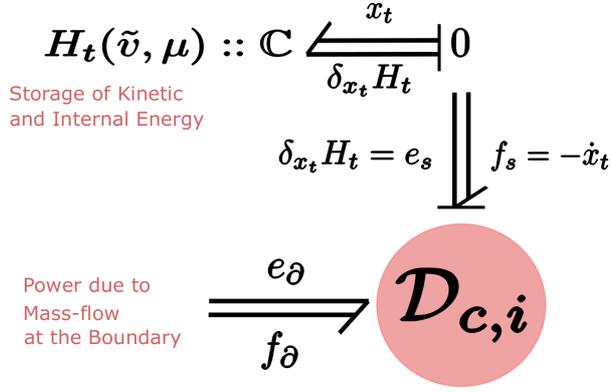}
\caption{Port-based representation of Euler Equation (\ref{eq:comp_isen_pH_eqns}) of compressible isentropic flow.}
\label{fig:Euler_isentropic}
\end{figure}

It is interesting to combine all the energy storage elements into one as well as combine all the power conserving elements to a new Stokes Dirac structure $\cl{D}_{c,i}$, as shown in Fig. \ref{fig:Euler_isentropic}.
The new storage element has its state variables $x_t = \stateV$ and its Hamiltonian $H_t$ given by the total energy of the system (the sum of kinetic and internal), i.e,
\begin{equation}\label{eq:Ham_total_isen}
H_t\stateV = H_k\stateV + H_i(\mu) = \int_M \half (*\mu) \vf \wedge * \vf + U(*\mu)\mu, 
\end{equation}
with flow and effort variables
\begin{equation}\label{eq:Ham_total_portVar}
\dot{x}_t = \TwoVec{\fVt}{\fMu}, \qquad \delta_{x_t} H_t = \TwoVec{\delta_{\vf} H_t}{\delta_{\mu} H_t} = \TwoVec{ \iota_v \mu}{\half  \iota_v \vf + h}.
\end{equation}
The new energy balance for $H_t$ is given by the following result.
\begin{proposition}\label{prop:Ht_dot_isen}
{ The rate of change of the total Hamiltonian $H_t$, given by (\ref{eq:Ham_total_isen}), along trajectories of its state variables $x_t = \stateV$ is expressed as}
\begin{equation}\label{eq:total_Ham_balance}
\dot{H}_t=  \int_{\Mbound} \eBoudT \wedge \fBoudT ,
\end{equation}
where the new boundary port variables $(\eBoudT,\fBoudT) \in \spKForm{0}{\Mbound} \times \spKForm{n-1}{\Mbound}$ are defined by
\begin{equation}\label{eq:total_port_boundary}
\eBoudT :=  \bound{\delta_{\mu} H_t} = \bound{(\half \iota_v \vf + h)}, \qquad \fBoudT := -\bound{\delta_{\vf} H_t}  = -\bound{(\iota_v \mu)}.
\end{equation}
\end{proposition} 
\begin{proof}
By starting from the energy balance for $\dot{H}_k$ in (\ref{eq:new_power_balance}) and using the equality (\ref{eq:power_balance_int_isent}), we have that
\begin{align*}
\dot{H}_k=&  \int_{\Mbound} \eBoud \wedge \fBoud   + \int_M e_d \wedge f_d
			= \int_{\Mbound} \bound{\half \iota_v\vf} \wedge -\bound{\iota_v\mu}   +\pair{\fstr}{\omega_v}_{\gothg}\\
			=& \int_{\Mbound} \bound{\half \iota_v\vf} \wedge -\bound{\iota_v\mu}   -\pair{\eMuI}{\fMu}_{V^*} - \int_{\Mbound} \eBoudInt\wedge\fBoudInt \\
			=& \int_{\Mbound} \bound{\half \iota_v\vf} \wedge -\bound{\iota_v\mu} - \dot{H}_i -  \int_{\Mbound} h\bound{} \wedge \bound{\iota_v\mu}\\
			=& \int_{\Mbound} \bound{(\half \iota_v\vf + h)} \wedge -\bound{\iota_v\mu} - \dot{H}_i.
\end{align*}
Thus, we have that 
$$\dot{H}_t= \dot{H}_k +  \dot{H}_i = \int_{\Mbound} \bound{(\half \iota_v\vf + h)} \wedge -\bound{\iota_v\mu},$$
which concludes the proof using (\ref{eq:total_port_boundary}).
\end{proof}
Physically the boundary effort variable $\eBoudT$ is known as the stagnation or total enthalpy at the boundary, while the boundary flow variable $\fBoudT$ represents the mass inflow through the boundary.
The power in the port $(\eBoudT,\fBoudT)$ represent the energy change due to the exchange of mass flow between the isentropic compressible flow system and its surroundings.

The overall Stokes-Dirac structure $\cl{D}_{c,i}$ for isentropic compressible flow that implements the power balance in (\ref{eq:total_Ham_balance}) is given by
\begin{equation}\label{eq:SDS_isentropic_system}
\begin{split}
\cl{D}_{c,i} = \{ &(f_s,\fBoudT,e_s,\eBoudT) \in \cl{B}_{c,i} | \\
			 \TwoVec{f_{\vf}}{f_{\mu}} &= \TwoVec{\extd e_{\mu} +  \frac{1}{*\mu} \iota_{\hat{e}_{\vf}} \extd \vf  }{\extd e_{\vf} },\\
			 \TwoVec{\eBoudT}{\fBoudT}  &=  \TwoTwoMat{0}{1}{-1}{0}\TwoVec{\bound{e_{\vf}}}{\bound{e_{\mu}}} \},
\end{split}
\end{equation}
where the total storage port variables are given by $f_s = ({f_\vf},{f_\mu}) \in \gothsStar $ and $e_{s} = (e_{\vf},e_{\mu}) \in \goths$.
The bond-space is now given by $\cl{B}_{c,i}= \cl{F}_{c,i}\times \cl{E}_{c,i}$, with the flow space $\cl{F}_{c,i}=  \spKFormM{1} \times \spKFormM{n} \times \spKForm{0}{\partial M}$ and the effort space $\cl{E}_{c,i} = \spKFormM{n-1} \times \spKFormM{0}  \times \spKForm{n-1}{\partial M}$.

{ The port-Hamiltonian dynamics for compressible isentropic flow is then recovered by setting $((-\fVt,-\fMu),\fBoudT,(\delta_{\vf} H_t,\delta_{\mu} H_t),\eBoudT) \in \cl{D}_{c,i},$ which yields
\begin{align}
\TwoVec{\fVt}{\fMu} =& \TwoVec{-\extd (\delta_{\mu} H_t) -  \iota_{v} \extd \vf } {-\extd (\delta_{\vf} H_t)},\label{eq:comp_isen_pH_eqns}\\
H_t(x_t) =& H_t\stateV = \int_M \half (*\mu) \vf \wedge * \vf + U(*\mu)\mu.
\end{align}
where the variational derivatives of $H_t$ are given by (\ref{eq:Ham_total_portVar}), and the boundary conditions are specified by the boundary port-variables $(\eBoudT,\fBoudT)$ given by (\ref{eq:total_port_boundary}).}

Finally we conclude by some remarks about the Dirac structure (\ref{eq:SDS_isentropic_system}) derived in this section.
First, this is exactly the Dirac structure which was just defined as a fundamental object in \citep{van2002hamiltonian}.
Here the geometrical structure that underpins this object has been rigorously explicated.

Second, the Dirac structure $\cl{D}_{c,i}$ given by (\ref{eq:SDS_isentropic_system}) is modulated by the state variables $\stateV$.
An interesting case occurs when the 2-form $\extd \vf =: \omega \in \spKFormM{2}$ is zero $\forall t$. In such case, the term $\iota_v\extd \tilde{v}$ in (\ref{eq:SDS_isentropic_system}) vanishes and the Dirac structure becomes a constant one in the bond space $\cl{B}_{c,i}$.
The 2-form $\omega$ is known as the \textit{vorticity form} which is also advected with the flow in ideal fluid flow \citep[Pg. 596]{Abraham1988}. Therefore, if the vorticity form is zero at $t=0$, it remains zero for all $ t>0$. Such type of fluid flow is called \textit{irrotational flow}\footnote{In \cite{van2002hamiltonian} it is erroneously remarked that the term $\iota_v\extd \tilde{v}$ also vanishes in two-dimensional flow, which is not the case.}.

Third, compared to the Dirac structure of the kinetic subsystem in (\ref{eq:SDS_kinetic_system_new}), the overall Dirac structure (\ref{eq:SDS_isentropic_system}) is exactly the same (if we exclude the distributed port).
This equivalence is due to the fact that both systems have the same state variables $\stateV$, but only differ in the Hamiltonian function which is independent from the underlying structure of the system.
This underlying structure composed of the external boundary port variables combined with the Lie-Poisson structure which governs the evolution equations of $\stateV$ independent of the Hamiltonian energy function.

\subsection{Adiabatic Compressible Flow}
\newcommand{\eSI}{\delta_{s} H_i}
\newcommand{\fS}{\dot{s}}
\newcommand{\Dad}{\cl{D}_{ad}}

Following the same line of thought as for the isentropic case, we can also extend the kinetic energy port-Hamiltonian system using the distributed port $(e_d,f_d)$ to model adiabatic flow.
The exact same procedure is applied for the internal energy storage element but for the extended state variable $x_i = (\mu,s)\in \cl{X}_i = V^*$.
Both the energy variables $(\mu,s)$ are advected quantities of the fluid. Thus the space of advected quantities in this case is $\bar{V}^* = \spKFormM{n}\times \spKFormM{0}$.

The internal energy Hamiltonian $\map{H_i}{\cl{X}_i}{\bb{R}}$ is now given by
\begin{equation}
H_i(\mu,s) = \int_M U(*\mu,s) \mu,
\end{equation}
where the specific internal energy $U(*\mu,s) = U(\rho,s)$ depends now on entropy as well.

The effort and flow variables of the internal energy storage element are given by
\begin{equation}\label{eq:port_variables_internal_adiabatic}
\begin{split}
\delta_{x_i}H_i &= \TwoVec{\eMuI}{\eSI} \in T_{x_i}^* \cl{X}_i \cong \bar{V} = \spKFormM{0}\times \spKFormM{n},\\
\dot{x}_i &= \TwoVec{\fMu}{\fS} \in T_{x_i} \cl{X}_i \cong \bar{V}^* = \spKFormM{n}\times \spKFormM{0}.
\end{split}
\end{equation}
The variational derivative of $H_i$ with respect to $\mu$ is given by Prop. \ref{prop:varHi_mu} while the variational derivative of $H_i$ with respect to $s$ is given by
\begin{equation}
\eSI = \parU{s}\mu = T\mu,
\end{equation}
using (\ref{eq:components_one_form_dU}).
The internal energy $H_i$ satisfies now the power balance
\begin{equation}
\dot{H}_i = \pair{\delta_{x_i}H_i}{\dot{x}_i}_{\bar{V}^*}= \pair{\eMuI}{\fMu}_{V^*} +  \pair{\eSI}{\fS}_{V^*}.
\end{equation}

\begin{figure}
\centering
\includegraphics[width=\textwidth]{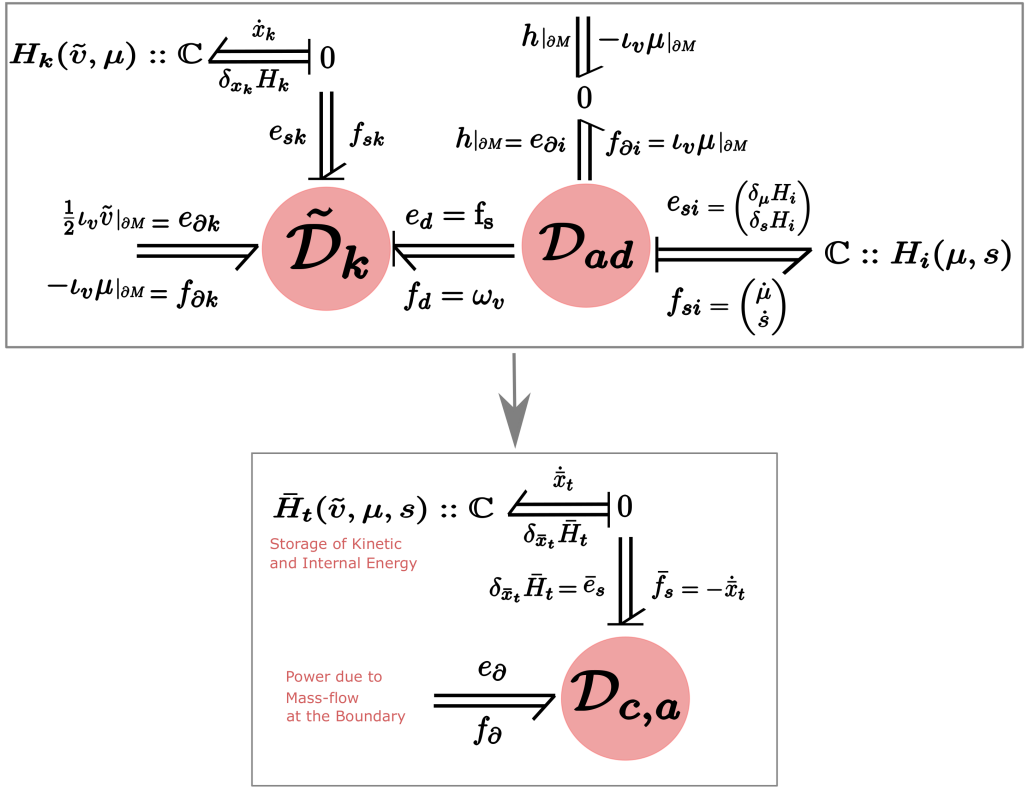}
\caption{Compressible adiabatic flow on a general manifold with permeable boundary. Top figure shows how to augment the kinetic energy system (\ref{eq:pH_kinetic_sys_new_1}) with the storage of internal energy, while the bottom figure shows a compact model with a combined storage element, Dirac structure, and boundary port.}
\label{fig:Euler_adiabatic}
\end{figure}

With reference to Fig. \ref{fig:Euler_adiabatic}, the { Dirac structure $\Dad$ used for} connecting the internal energy port $(\delta_{x_i}H_i,\dot{x}_i)$ to the distributed port $(e_d,f_d)$ is given by
{
\begin{equation}
\begin{split}
\Dad  = \{ (\fSint, &\fBoudInt, f_d,\eSint,\eBoudInt,e_d) \in \cl{B}_{ad}  | \\
			 \TwoVec{e_d}{\fSint} &= \TwoTwoMat{0}{\mapTdual{(\mu,s)} }{-\mapT{(\mu,s)}}{0} \TwoVec{f_d}{\eSint},\\
			 \TwoVec{\eBoudInt}{\fBoudInt}  &= \begin{pmatrix}
			 1 	& 	0	& 0 \\
			 0  & 	0 	& \bound{*\mu}
			 \end{pmatrix}
			 \ThrVec{\bound{e_\mu}}{\bound{e_s}}{\bound{f_d}}\},\label{eq:Adb_Dirac_structure}
\end{split}
\end{equation}
where $\fSint := (f_\mu,f_s) \in \bar{V}^*  = \spKFormM{n}\times \spKFormM{0}$ and $\eSint := (e_\mu,e_s) \in \bar{V} = \spKFormM{0}\times \spKFormM{n}$, $(\fBoudInt,\eBoudInt) \in \spKForm{0}{\partial M}\times \spKForm{n-1}{\partial M}$, and $(e_d,f_d) \in \spKFormM{1}\times \spKFormM{n-1}$. The bond-space $ \cl{B}_{ad}$ is then given by the product of the aforementioned spaces of forms.}

The map $\map{\mapT{(\mu,s)}}{\gothg}{\bar{V}^*}$ and its dual $\map{\mapTdual{(\mu,s)}}{\bar{V}}{\gothgstar}$ are defined, respectively, for any $\omega \in \gothg$ and $(e_\mu,e_s) \in \bar{V}$ as
\begin{equation}
\mapT{(\mu,s)}(\omega) := \TwoVec{\mapT{\mu}(\omega)}{\mapT{s}(\omega)}, \qquad \mapTdual{(\mu,s)}(e_\mu,e_s) = \mapTdual{\mu}(e_{\mu}) + \mapTdual{s}(e_s),
\end{equation}
which allows one to rewrite { the first equation in (\ref{eq:Adb_Dirac_structure})} as
\begin{equation}\label{eq:MTF_mu_s_2}
\ThrVec{e_d}{f_\mu}{f_s} = 
\begin{pmatrix}
0 	& 	\mapTdual{\mu}	& \mapTdual{s} \\
-\mapT{\mu}  	& 0 	& 0 \\
-\mapT{s}  	& 0 	& 0
\end{pmatrix}
\ThrVec{f_d}{e_\mu}{e_s}.
\end{equation}
For the choice of $\mu$ and $s$ as the advected parameters, the maps $\mapT{\mu},\mapT{s}$ and $\mapTdual{\mu},\mapTdual{s}$ are given in Table \ref{table:intercon_maps}.

{ 
The power balance that the Dirac structure $\Dad$ encodes is given by the following result.
\begin{proposition}\label{prop:Dad_power_balance}
The Dirac structure $\Dad$ given by (\ref{eq:Adb_Dirac_structure}) is a power continuous structure, such that 
\begin{equation}\label{eq:D_ad_equality}
\pair{e_d}{f_d}_\gothg +  \pair{\eSint}{\fSint}_{\bar{V}^*} + \int_{\Mbound} \eBoudInt\wedge\fBoudInt =0.
\end{equation}
\end{proposition}
\begin{proof}
Using (\ref{eq:Adb_Dirac_structure} - \ref{eq:MTF_mu_s_2}), we have that
\begin{align*}
\pair{e_d}{f_d}_\gothg 	&= \pair{\mapTdual{\mu}(e_\mu)}{f_d}_\gothg + \pair{\mapTdual{s}(e_s)}{f_d}_\gothg \\
 						&= \pair{e_\mu}{\mapT{\mu}(f_d)}_{V^*} + \int_{\Mbound} \surfPhiDTA{f_d}(\mu,e_\mu) 
 						 + \pair{e_s}{\mapT{s}(f_d)}_{V^*} + \int_{\Mbound} \surfPhiDTA{f_d}(s,e_s)\\
 						&= \pair{e_\mu}{-f_\mu}_{V^*} - \int_{\Mbound} \bound{(*\mu f_d)} \wedge \bound{e_\mu} +  \pair{e_s}{-f_s}_{V^*} + 0,\\
 						&= - \pair{\eSint}{\fSint}_{\bar{V}^*} - \int_{\Mbound} \eBoudInt\wedge\fBoudInt,
\end{align*}
which follows from (\ref{eq:pair_Rmap}) and the interconnection map expressions (and their corresponding surface terms) in Table \ref{table:intercon_maps}.
%
\end{proof}

With reference to Fig. \ref{fig:Euler_adiabatic}, the Dirac structure $\Dad$ is used to model adiabatic compressible flow by setting its ports to 
$$((\fMu,\fS),\bound{\iota_v \mu},\omega_v, (\eMuI,\eSI),\bound{h},\fstr)\in\Dad.$$
}
Therefore, following exactly the steps shown in (\ref{eq:fMu}), the evolution of $s$ is given by
\begin{equation}\label{eq:fS}
\dot{s} = -\mapT{s}(\omega_v) = -\LieD{v}{s},
\end{equation}
and the evolution of $\mu$ is the same as the isentropic case in (\ref{eq:fMu}).
Moreover, using (\ref{eq:fbody}) and the definition of $\mapTdual{s}$ we have that
\begin{equation}\label{eq:fbody_adiabatic}
\begin{split}
\fstr &= \mapTdual{\mu}(\eMuI)+\mapTdual{s}(\eSI) 
	   = - (*\mu)\extd h +  *(\eSI) \extd s \\
	   &= - (*\mu)\extd h +  *(T\mu) \extd s = - (*\mu)\extd h + T (*\mu)\extd s = -\extd p,
\end{split}
\end{equation}
where the fourth equality follows from the commutativity of the Hodge star with functions, and the final results follows from Gibbs equation (\ref{eq:Gibbs_eqn_2}).
Therefore, both (\ref{eq:fS}) and (\ref{eq:fMu}) correctly represent the evolution of the entropy $s$ and the mass form $\mu$ as being advected with the flow, while (\ref{eq:fbody_adiabatic}) correctly represents the stress forces due to pressure consistent with the thermodynamics of the system.

\newcommand{\Ht}{\bar{H}_t}
\newcommand{\xt}{\bar{x}_t}
Finally we conclude by a more compact port-Hamiltonian model for adiabatic compressible flow, as shown in Fig. \ref{fig:Euler_adiabatic}.
The new storage element has its state variables $\xt := (\vf,\mu,s)$ and the total Hamiltonian $\Ht$ given by
\begin{equation}\label{eq:Ham_total_ad}
\Ht(\vf,\mu,s) = \int_M \half (*\mu) \vf \wedge * \vf + U(*\mu,s)\mu, 
\end{equation}
with flow and effort variables
\begin{equation}\label{eq:Ht_varDeriv}
\dot{\bar{x}}_t = \ThrVec{\fVt}{\fMu}{\fS}, \qquad \delta_{\xt} \Ht = \ThrVec{\delta_{\vf} \Ht}{\delta_{\mu} \Ht}{{\delta_{s} \Ht}} = \ThrVec{ \iota_v \mu}{\half  \iota_v \vf + h}{T\mu}.
\end{equation}
Interestingly, the new energy balance for $\Ht$ is given by the same power balance as for the isentropic case as will be proven in the following.
\begin{proposition}
{ The rate of change of the total Hamiltonian $\Ht$, given by (\ref{eq:Ham_total_ad}), along trajectories of its state variables $\xt = (\vf,\mu,s)$ is expressed as}
\begin{equation}\label{eq:total_Ham_balance2}
\dot{\bar{H}}_t=  \int_{\Mbound} \eBoudT \wedge \fBoudT ,
\end{equation}
where the same boundary port variables defined before in (\ref{eq:total_port_boundary}).
\end{proposition} 
\begin{proof}

The proof follows exactly the one of Prop. \ref{prop:Ht_dot_isen}, where the pairing $\pair{e_d}{f_d}_{\gothg} = \pair{\fstr}{\omega_v}_{\gothg}$ is substituted by the power balance given by Prop. \ref{prop:Dad_power_balance}.
\end{proof}

\begin{remark}

The reason why the energy balance (\ref{eq:total_Ham_balance2}) for adiabatic flow is equivalent to the one for isentropic flow in (\ref{eq:total_Ham_balance}) is mainly due to the vanishing of the surface term $\surfPhiDTA{f_d}(s,e_s)$ in the proof of Prop. \ref{prop:Dad_power_balance}, which follows from Table \ref{table:intercon_maps}.
The physical intuition behind this observation is the fact that adiabatic flow corresponds to conservation of entropy due to no exchange of heat with the surroundings. Thus, it is natural that no increase of internal energy occurs due to heat exchange through the boundary, and the only way for internal energy to increase is due to mass flow through the boundary.
\end{remark}

The overall Stokes-Dirac structure $\cl{D}_{c,a}$ for adiabatic compressible flow that implements the power balance in (\ref{eq:total_Ham_balance2}) is given by
{
\begin{equation}\label{eq:SDS_adiabatic_system}
\begin{split}
\cl{D}_{c,a} = \{ &(\bar{e}_s,\bar{f}_s,\eBoudT,\fBoudT) \in \cl{B}_{c,a} | \\
			 \ThrVec{f_{\vf}}{f_{\mu}}{f_{s}} &= \ThrVec{\extd e_{\mu} +  \frac{1}{*\mu} \iota_{\hat{e}_{\vf}} \extd \vf - \frac{* e_s}{*\mu}\extd s }{\extd e_{\vf} }{\frac{1}{*\mu} \iota_{\hat{e}_{\vf}} \extd s},\\
			 \TwoVec{\eBoudT}{\fBoudT}  &= 
			 \begin{pmatrix}
			 0 & 1 & 0\\ -1 & 0 & 0
			 \end{pmatrix}
			 \ThrVec{\bound{e_{\vf}}}{\bound{e_{\mu}}}{\bound{e_{s}}} \},
\end{split}
\end{equation}
{
where the total storage port variables are given by $\bar{f}_s = ({f_\vf},{f_\mu},f_s) \in \spKFormM{1} \times \spKFormM{n} \times \spKFormM{0}$ and $\bar{e}_{s} = (e_{\vf},e_{\mu},e_s) \in \spKFormM{n-1} \times \spKFormM{0} \times \spKFormM{n}$.
The bond-space is given by $\cl{B}_{c,a}= \cl{F}_{c,a}\times \cl{E}_{c,a}$, with the flow space $\cl{F}_{c,a}=  \spKFormM{1} \times \spKFormM{n} \times \spKFormM{0} \times \spKForm{0}{\partial M}$ and the effort space $\cl{E}_{c,a} = \spKFormM{n-1} \times \spKFormM{0} \times \spKFormM{n}  \times \spKForm{n-1}{\partial M}$.

Finally, the port-Hamiltonian dynamics for compressible adiabatic flow is then recovered by setting $((-\fVt,-\fMu,-\fS),\fBoudT,(\delta_\vf \Ht,\delta_{\mu} \Ht,\delta_{s} \Ht),\eBoudT) \in \cl{D}_{c,a},$ which yields
\begin{align}
\ThrVec{\fVt}{\fMu}{\fS} =& \ThrVec{-\extd (\delta_{\mu} \Ht) -  \iota_{v} \extd \vf  + (* \delta_{s} \Ht/*\mu)\extd s} {-\extd (\delta_{\vf} \Ht)}{-\iota_v \extd s},\\
\Ht(\xt) =& \Ht(\vf,\mu,s) = \int_M \half (*\mu) \vf \wedge * \vf + U(*\mu,s)\mu, 
\end{align}
where the following equality was used
\begin{equation}
-\dot{s} = \LieD{v}{s}= \extd \iota_v s +  \iota_v \extd s = \iota_v \extd s = \frac{1}{*\mu}\iota_{\hat{e}_{\vf}} \extd s.
\end{equation}
The variational derivatives of $\Ht$ are given by (\ref{eq:Ht_varDeriv}), and the boundary conditions are specified by the boundary port-variables $(\eBoudT,\fBoudT)$ given by (\ref{eq:total_port_boundary}).

}

\section{Port-Hamiltonian Modeling of Incompressible Flow}
\newcommand{\Dinc}{\cl{D}_{inc}}
\newcommand{\eBoudC}{{e}_{\partial c}}
\newcommand{\fBoudC}{{f}_{\partial c}}

\subsection{Conservation of Volume}
In the physical world, it is observed from experiments that the compressibility of a fluid could be neglected when the speed of a body within the fluid is much lower than the speed of sound.
In this case, the flow is approximated to be incompressible which is characterized mathematically by the conservation of the volume form $g_t^* \volF = \volF$.

Let the top-form given by $g_t^* \volF$ have a density $J(g_t) \in \spFn{M}$ defined such that $g_t^* \volF =J(g_t) \volF $.
The incompressibility  condition implies that $J(g_t) = 1$ for all times and at all points $x\in M$.

By the Lie derivative rule, 
$
\dt (g_t^* \volF) = g_t^*(\LieD{v}{\volF}),
$
an equivalent condition for incompressible flow is $\LieD{v}{\volF} = 0$.
Therefore, in incompressible flow the time-independent volume form is also an advected quantity, or more correctly it is frozen in the fluid.

Using properties of the Lie derivative, one also has that
$
\LieD{v}{\volF} = \divr{v} \volF = 0 \implies \divr{v} = 0,
$
as well as
$
\LieD{v}{\volF} = \extd \iota_v \volF = \extd \omega_v=  0.
$
Therefore, the following are all equivalent conditions for incompressible flow:
\renewcommand{\labelenumi}{\roman{enumi})}
\begin{enumerate*}
\item $ J(g_t) = 1, \qquad$
\item $\LieD{v}{\volF} = 0,\qquad$
\item $\divr{v} = 0,\qquad$
\item $ \extd \omega_v = 0$.
\end{enumerate*}

In the case of incompressible flow, the aforementioned conditions restricts the configuration space of the fluid flow to a subgroup of $\diffG$ defined by
\begin{equation}
\cl{D}_\text{vol}(M) := \{ g\in \diffG | J(g) =1\},
\end{equation}
This subgroup is known in the literature as the volume-preserving diffeomorphism group, which was shown in \cite{ebin1970groups} to be a Lie-subgroup of $\diffG$.
The corresponding Lie sub-algebra is given by the divergence-free vector fields $\spVecX{M}{\text{vol}}$ defined as
$
\spVecX{M}{\text{vol}} := \{ v\in \spVecM | \divr{v} = 0\}.
$
Condition (iv) also implies that the differential form representation of the Lie algebra $\spVecX{M}{\text{vol}}$ corresponds to the closed $n-1$ forms.

\subsection{Port-Hamiltonian Model}

In incompressible flow, the fluid is characterized only by kinetic energy and no internal energy is present.
The pressure function in incompressible flow no longer has its thermodynamic nature as in compressible flow, but rather acts as a Lagrange multiplier that enforces the incompressiblity of the flow.
Therefore, an incompressible flow system is classified as a constrained mechanical system, not a thermodynamic system.

To represent incompressible flow in the port-Hamiltonian framework, the kinetic energy subsystem (\ref{eq:pH_kinetic_sys_new_1}-\ref{eq:pH_kinetic_Ham_new}) already contains all the energy stored, its energy variables, and its corresponding interconnection structure.
The difference now is that the system (\ref{eq:pH_kinetic_sys_new_1}) no longer has the state space $\cl{X}= \gothsStar = \gothgstar\times V^* = \spKFormM{1}\times \spKFormM{n},$
but instead the constrained state space $\cl{X}_c$ defined by
\begin{equation}
\cl{X}_c := \gothgstar_c\times V^* = \tilde{C}^1(M)\times \spKFormM{n},
\end{equation}
where $\gothgstar_c := \tilde{C}^1(M)\subset \spKFormM{1}$ denotes the space of co-closed 1-forms defined by
$\tilde{C}^1(M):=\{\vf\in \spKFormM{1} | \extd * \vf = 0\}.$
The incompressiblity constraint $\extd * \vf = \extd \omega_v =  0$ is equivalent to the conservation of the volume form, as discussed in the previous section.

For the port-Hamiltonian system (\ref{eq:pH_kinetic_sys_new_1}) to correctly represent incompressible flow, the distributed port $(e_d,f_d) = (\fstr,\omega_v)$ needs to be adapted to model stress forces that impose the incompressiblity constraint.
Following the exact same manner as in the previous section, the key point that allows building the port-based model of incompressible flow is that the volume form $\volF$ is frozen in the fluid (i.e. an advected quantity).

With reference to Fig. \ref{fig:Euler_incompr}, the interconnection to model incompressible flow is achieved by the Dirac structure $\Dinc$ given by
\begin{equation}
\begin{split}
\Dinc  = \{ (\fBoudInt, &f_{di},\eBoudInt,e_{di}) \in \cl{B}_{inc}  | \\
			 \TwoVec{e_{di}}{0} &= \TwoTwoMat{0}{\mapTdual{\volF} }{-\mapT{\volF}}{0} \TwoVec{f_d}{p},\\
			 \TwoVec{\eBoudInt}{\fBoudInt}  &= \TwoTwoMat{1}{0}{0}{1}\TwoVec{\bound{p}}{\bound{f_{di}}}\},\label{eq:Int_Dirac_structure_Incom}
\end{split}
\end{equation}
where the bond-space $\cl{B}_{inc}= \cl{F}_{inc}\times \cl{E}_{inc}$ is the product space of the flow space $\cl{F}_{inc}= \spKForm{n-1}{\partial M} \times \spKForm{n-1}{M} $ and the effort space $\cl{E}_{inc} = \spKForm{0}{\partial M} \times \spKForm{1}{M}$.

The power continuity of $\Dinc$ is given by the following result.
\begin{proposition}\label{prop:Dinc_power_balance}
The Dirac structure $\Dinc$ given by (\ref{eq:Int_Dirac_structure_Incom}) is a power continuous structure, such that 
\begin{equation}\label{eq:D_inc_equality}
\pair{e_{di}}{f_{di}}_\gothg +  \int_{\Mbound} \eBoudInt\wedge\fBoudInt =0.
\end{equation}
\end{proposition}
\begin{proof}
Using (\ref{eq:pair_Rmap}) and (\ref{eq:Int_Dirac_structure_Incom}), we have that
\begin{align*}
\pair{e_{di}}{f_{di}}_\gothg &=  \pair{\mapTdual{\volF}(p)}{f_d}_\gothg
 						= \pair{p}{\mapT{\volF}(f_d)}_{V^*} + \int_{\Mbound} \surfPhiDTA{f_d}(\volF,p) \\
 						&= \pair{p}{0}_{V^*} - \int_{\Mbound} \bound{f_d} \wedge \bound{p}
 						= - \int_{\Mbound} \eBoudInt\wedge\fBoudInt,
\end{align*}
which follows from the definition of $\mapTdual{\volF}$ and $\surfPhiDTA{f_d}$ given in Table \ref{table:intercon_maps} with $\volF$ instead of $\mu$ as the advected quantity.
\end{proof}

\begin{figure}
\centering
\includegraphics[width=\textwidth]{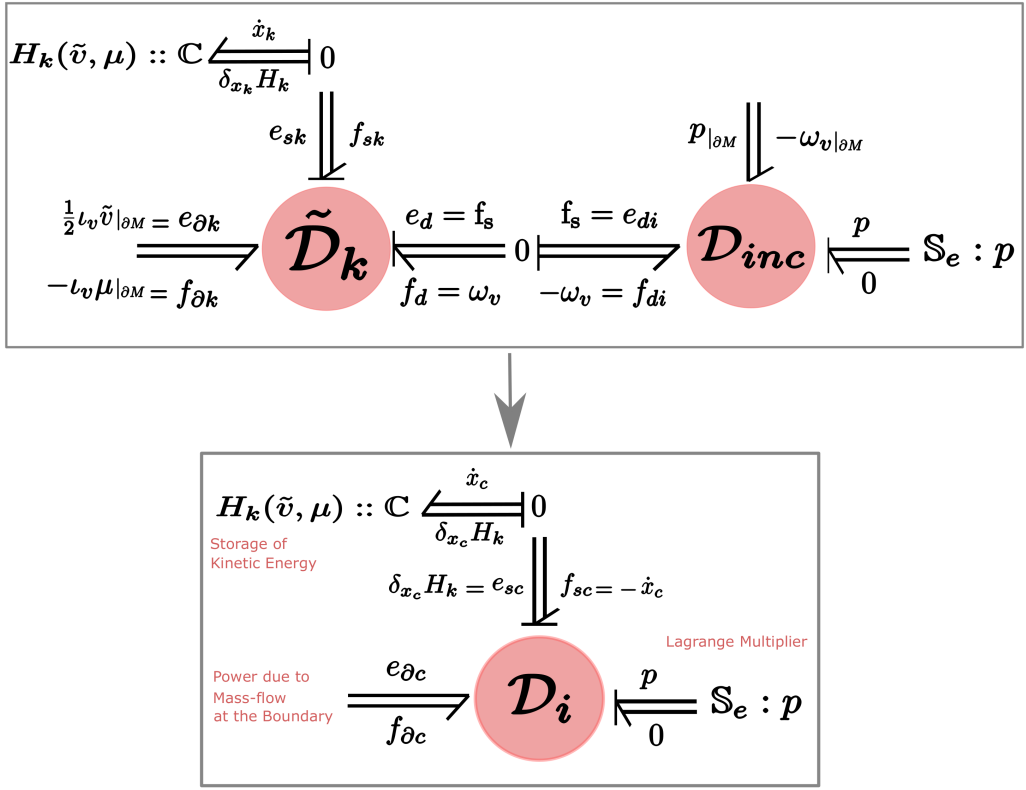}
\caption{Incompressible inhomogeneous flow on a general manifold with permeable boundary. Top figure shows how to augment the kinetic energy system (\ref{eq:pH_kinetic_sys_new_1}) with the pressure as a Lagrange multiplier, while the bottom figure shows a compact model with a combined storage element, Dirac structure, boundary port, and constraint distributed port.}
\label{fig:Euler_incompr}
\end{figure}

\begin{remark}
A special feature of the Dirac structure $\Dinc$ is that it has three ports, one of which, namely $(p,0)$, does not affect the power balance (\ref{eq:D_inc_equality}).
The power flowing through the port $(p,0)$ is always zero such that the pressure function acts as a Lagrange multiplier enforcing the incompressiblity constraint.
A direct consequence is that the Dirac structure $\Dinc$ is defined as a subspace on the two ports $(\eBoudInt,\fBoudInt)$ and $(e_{di}, f_{di})$ only, as shown in (\ref{eq:Int_Dirac_structure_Incom}). 
\end{remark}

The Dirac structure (\ref{eq:Int_Dirac_structure_Incom}) is then used to model incompressible flow by setting its ports to
$(\bound{\iota_v \mu},-\omega_v, \bound{p},\fstr)\in\Dinc.$
Therefore from (\ref{eq:Int_Dirac_structure_Incom}) it follows that
\begin{equation}\label{eq:incom_const}
0 = -\mapT{\volF}(-\omega_v) = \LieD{v}{\volF} = \extd \iota_v \volF =\extd * \vf,
\end{equation} 
and similarly
\begin{equation}\label{eq:press_forces}
\fstr = \mapTdual{\volF}(p) = -\extd p.
\end{equation} 
Thus, both the incompressiblity constraint and the forces due to pressure are properly modeled in (\ref{eq:incom_const}) and (\ref{eq:press_forces}), respectively.

\begin{remark}
From the power balance (\ref{eq:D_inc_equality}) it is worth noticing that if we neglect the surface term, the power flow through distributed port $(e_d,f_d) = (\fstr,\omega_v)$ is equal to zero, which is a consequence that the work done due to the pressure $\pair{p}{0}_{V^*}$ is equal to zero.
The pressure $p$ acts as a Lagrange multiplier that only enforces the incompressiblity constraint, and no longer has its thermodynamic nature in incompressible flow, which is considered a limit case of the general compressible flow.
\end{remark}

To summarize, the explicit port-Hamiltonian dynamical model of (inhomogeneous) incompressible flow in terms of the constrained state variable $x_c: = \stateV \in \cl{X}_c = \gothgstar_c\times V^*$ is given by
\begin{align}
\TwoVec{\fVt}{\fMu} =& \TwoVec{-\extd (\eMu) -  \iota_{v} \extd \vf } {-\extd (\eVt)} - \TwoVec{ \frac{1}{*\mu} \circ \extd}{0}p, \label{eq:pH_inc_1}\\
0 =& \begin{pmatrix}  \extd \circ \frac{1}{*\mu} &  0\end{pmatrix} \TwoVec{\eVt}{\eMu}, \label{eq:pH_inc_2}\\
H_k(x_c) =& H_k\stateV = \int_M \half (*\mu) \vf \wedge * \vf.\label{eq:pH_kinetic_Ham_inc}
\end{align}
where the variational derivatives are given by (\ref{eq:varDeriv_stateV}), and the pressure function $p\in \spFn{M}$ is a distributed Lagrange multiplier.
The energy balance for $H_k$ is given by the following result.
\begin{proposition}\label{prop:Hk_dot_inc}
The rate of change of the total Hamiltonian $H_k$, given by (\ref{eq:pH_kinetic_Ham_inc}), along trajectories of (\ref{eq:pH_inc_1}) is expressed as
\begin{equation}\label{eq:Ham_balance_H_k_inc}
\dot{H}_k=  \int_{\Mbound} \eBoudC \wedge \fBoudC ,
\end{equation}
where the boundary port variables $\eBoudC,\fBoudC \in \spKForm{0}{\Mbound} \times \spKForm{n-1}{\Mbound}$ are defined by
\begin{equation}\label{eq:port_boundary_inc}
\begin{split}
\eBoudC :=&  \bound{\delta_{\mu} H_k} + \bound{\left(\frac{p}{*\mu}\right)} = \bound{\left(\half \iota_v \vf + \frac{p}{*\mu}\right)},\\
\fBoudC :=& -\bound{\delta_{\vf} H_k}  = -\bound{(\iota_v \mu)}.
\end{split}
\end{equation}
\end{proposition} 
\begin{proof}
By starting from the energy balance for $\dot{H}_k$ in (\ref{eq:new_power_balance}) and using the equality (\ref{eq:D_inc_equality}) with $\pair{e_{di}}{f_{di}}_\gothg = \pair{\fstr}{-\omega_v}_\gothg$ , we have that
\begin{align*}
\dot{H}_k=&  \int_{\Mbound} \eBoud \wedge \fBoud   + \int_M e_d \wedge f_d
			= \int_{\Mbound} \bound{\half \iota_v\vf} \wedge -\bound{\iota_v\mu}   -\pair{\fstr}{-\omega_v}_{\gothg}\\
			=& \int_{\Mbound} \bound{\half \iota_v\vf} \wedge -\bound{\iota_v\mu}  + \int_{\Mbound} \eBoudInt\wedge\fBoudInt \\
			=& \int_{\Mbound} \bound{\half \iota_v\vf} \wedge -\bound{\iota_v\mu}  +  \int_{\Mbound} \bound{p} \wedge \bound{-\omega_v}\\
			=& \int_{\Mbound} \bound{\half \iota_v\vf} \wedge -\bound{\iota_v\mu}  +  \int_{\Mbound} \bound{\left(\frac{p}{*\mu}\right)} \wedge \bound{-\iota_v\mu}\\
			=& \int_{\Mbound} \bound{\left(\half \iota_v\vf + \frac{p}{*\mu}\right)} \wedge -\bound{\iota_v\mu}.
\end{align*}
\end{proof}

Finally, the Stokes-Dirac structure $\cl{D}_i$ that encodes the power balance (\ref{eq:Ham_balance_H_k_inc}) is given by
\begin{equation}\label{eq:SDS_inc}
\begin{split}
\cl{D}_{i} = \{ &(f_{sc},\fBoudC,e_{sc},\eBoudC) \in \cl{B}_{i} | \\
			 \TwoVec{f_{\vf}}{f_{\mu}} &= \TwoVec{\extd e_{\mu} +  \frac{1}{*\mu} \iota_{\hat{e}_{\vf}} \extd \vf }{\extd e_{\vf}} + \TwoVec{ \frac{1}{*\mu} \circ \extd}{0}p,\\
			 0 =& \begin{pmatrix}   \extd \circ \frac{1}{*\mu} &  0\end{pmatrix} \TwoVec{e_\vf}{e_\mu},\\
			 \TwoVec{\eBoudC}{\fBoudC}  &= \TwoTwoMat{0}{1}{-1}{0}\TwoVec{\bound{e_{\vf}}}{\bound{e_{\mu}}} + \TwoVec{ \frac{1}{*\mu}}{0} \bound{p} \},
\end{split}
\end{equation}
where the boundary port variables $\eBoudC,\fBoudC \in \spKForm{0}{\Mbound} \times \spKForm{n-1}{\Mbound}$ and total storage port variables are given by $f_{sc} = ({f_\vf},{f_\mu}) \in  \tilde{C}^1(M)\times \spKFormM{n}$ and $e_{sc} = (e_{\vf},e_{\mu}) \in C^{n-1}(M)\times \spKFormM{0}$, where $ C^{n-1}(M) \subset \spKFormM{n-1}$ is the space of closed $n-1$ forms.
The bond-space $\cl{B}_{i}$ is given accordingly by the product of the aforementioned spaces, as usual.

\begin{remark}
For inhomogeneous incompressible flow, one can derive the Lie-Poisson part of the Dirac structure (\ref{eq:SDS_inc}) by semi-direct product reduction (\textit{cf.} \citep{marsden1976well}) starting from the configuration space $\cl{D}_\text{vol}(M) \ltimes V$, where $\cl{D}_\text{vol}(M)$ represents the volume preserving diffeomorphisms on $M$.
However, in the modular approach we presented, a re-derivation of the underlying structure is unnecessary as the open ports of the system (\ref{eq:pH_kinetic_Ham_new}) were used to constraint the state space to the dual of the Lie algebra of $\cl{D}_\text{vol}(M) \ltimes V$.

For the case of homogeneous incompressible flow, one no longer has the semi-direct product structure as the mass form becomes constant in space and is no longer advected.
In this case, the standard Hamiltonian reduction theorems can be used to derive the Lie-Poisson structure as in \citep{ebin1970groups,Arnold1992,Modin2011}.
\end{remark}

\section{Conclusion}
\begin{figure}
\centering
\includegraphics[width=\textwidth]{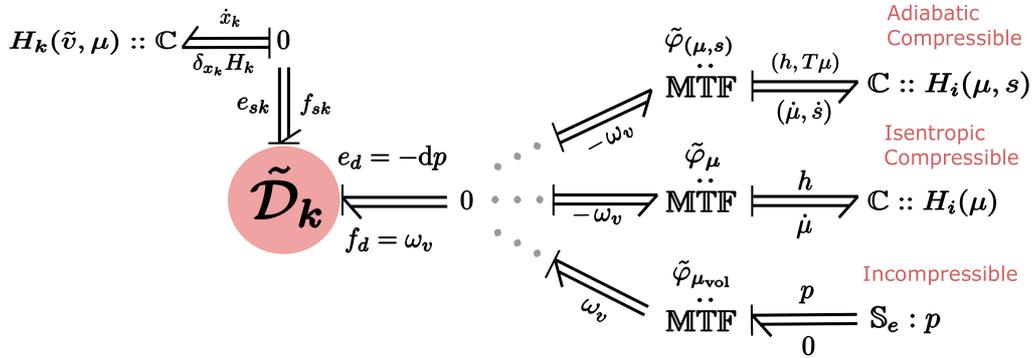}
\caption{Graphical representation of compressible and incompressible Euler Equations on a closed manifold ($\Mbound = \emptyset$) showing modularity of the port-Hamiltonian framework. For a general permeable manifold, the interconnection is achieved using Dirac structures.}
\label{fig:All_Euler}
\end{figure}

In this two-parts paper, a systematic procedure to model a variety of fluid dynamical systems on general Riemannian manifolds was presented.
The procedure was demonstrated for developing decomposed and open port-Hamiltonian models for (ideal) compressible and incompressible flow with variable boundary conditions.
The models presented are all geometric and thus are globally defined independently of a choice of coordinates on the spatial manifold $M$, thanks to the formulation of the equations using exterior calculus.

We have highlighted in this article series that the philosophy of port-Hamiltonian modeling is different from the classic approach of deriving Hamiltonian equations using variational principles \citep{Arnold1992,Marsden1972,marsden1984semidirect}.
The philosophy of the latter approach is a top-down procedure starting from the \textit{total} energy (Hamiltonian) defined on the cotangent bundle of the system's configuration space and then deriving the \textit{total} equations of motion governing the system.
Whereas, the philosophy of the port-Hamiltonian framework is a bottom-up procedure starting from subsystems that are interconnected together to form the complex total system.
The straightforward advantage compared to the variational approach, is that simply the model is updated by adding a new subsystem without re-deriving the whole dynamical equations.
This has been demonstrated by extending the subsystem corresponding to kinetic energy storage to three different models, summarized in Fig. \ref{fig:All_Euler}.

One advantage of our presented decomposed models is that they are \textit{open models}.
Using the \textit{open boundary port} or an extra distributed port, the derived models can be extended to more complicated fluid systems with (potentially) other physical domains, like e.g. structural mechanics or electromagnetism.
The only constraint when coupling subsystems of different nature is that one finds the physical reason for why they can be coupled in the first place.
If systems of different complexity are to be coupled (such as a fluid and a structure) a physical condition must be present that effects the suitable complexity reduction of the ports of the more
complex system (such as the no-slip condition for coupling fluids and structures) so that they can be coupled.

Another advantage of our work is that our framework allows to decompose a fluid domain into several imaginary subdomains whose equally imaginary boundaries of course do not prevent the flow between these subdomains.
But since the thus constituted subsystems must now be connected through a Dirac structure that routes the energy, between adjoining domains, we obtain control over precisely this energy flow.
This promises an avenue to ensure compatibility of our energy-aware decomposition with correspondingly designed structure-preserving numerical schemes, e.g. \citep{Seslija2011,Seslija2012a} that choose to discretize some parts of a fluid's domain in a more refined fashion than others, as is often needed.

%

\section*{Funding}
This work was supported by the PortWings project funded by the European Research Council [Grant Agreement No. 787675]

\bibliographystyle{abbrv}
\bibliography{Paper_References}

\end{document}